# Atoms of None of the Elements Ionize, While Atoms of Inert Behavior Split by Photonic Current

**Mubarak Ali**

Islamabad, Park Road, Pakistan, mubarak3974@gmail.com, mubarak74@mail.com, mubarak60@hotmail.com, https://www.researchgate.net/profile/Mubarak-Ali-7, http://orcid.org/0000-0003-1612-6014

**Abstract:** Ion formation is a concept in science. The positive or negative charge is the subject of many studies. A charge forms in published studies when an atom loses or gains an electron. But ion formation in atoms challenges the related principles and phenomena. An electric current is also a source of all the experimental details and procedures. But an electric current is actually a photonic current. Notably, suitable atoms can execute the interstate dynamics of qualified electrons. Atoms also undergo transition states. Solid atoms can elongate or expand. Gaseous atoms can contract or shrink. Inert atoms can split into electron streams. Solid atoms can elongate. By carrying photons, when electron streams impinge on naturally elongated solid atoms, those atoms can further elongate. If those solid atoms do not elongate, they at least deform by distorting. The characteristics of the photons are apparent when the flowing inert gas atoms split under the excessive current field. The splitting of inert gas atoms by the current field indicates that an electric current is a photonic current. The carrying of photons by the electron streams also indicates that it is a photonic current. The photons lighting the air also reveal a photonic current. In the microscopic analysis of a material, the image formation on the computer screen is due to the resolving power of the featured photons. In a suitable material, the bandgap is not due to a conduction bandgap. It is due to the gaps between the electronic states of the bound atoms. The theoretical frameworks discussed here directly support the material investigations. The study also discusses several physical and chemical aspects of science.

## 1.0. Introduction

In atoms of different elements, it is customary to consider a negative or positive charge. To describe related physical and chemical aspects, it is common to consider ion formation. A negative or positive charge is obligatory when an atom gains or loses an electron, respectively. Such ions form the basis of a chemical or physical process. Ions are species that possess either a net negative or a net positive charge. The net negative charge is an anion. An anion would be attracted to the anode.

A cation is not like that. In the atom of any element, an anion forms due to the gain of an electron. In the published literature, it is also a negatively charged ion. A cation forms when an atom loses an electron. In the published literature, a cation is also a positively charged ion. So, a cation would be attracted to the cathode. The existing study does not agree with these details.

In the processing of salt solutions, Sir Svante August Arrhenius discussed the formation of Faraday ions and was awarded the Nobel Prize in 1903 [1]. For performing work on the state equation, Sir Johannes Diderik van der Waals was awarded the Nobel Prize in 1910 [2]. The study of van der Waals interactions has been a topic of intense debate. Despite this, these interactions only present a partial understanding of the subject. For work on the elementary charge of electricity and on the photoelectric effect, there was also the Nobel Prize [3]. An ion has a net electrical charge, as discussed elsewhere [4]. The ion pair consists of a free electron and a positive ion. A study elsewhere [5] discussed this. However, these investigations are not the case here.

In the processing or synthesis of different materials by any means, the voltage and current define a power, which determines the source of the electric current. The flow of electrons in the published processes and methods refers to the electric current. In some studies, the electrons are also considered charged particles. In material processing or material synthesis, every published study considers the electric current as an input source of power. It is not true. An electric current does not exist. There should be a photonic current.

The force exerted at the electron level determines the energy behavior of an amalgamating nanoparticle or particle in solution [6]. The development of shaped particles under varying concentrations of gold precursors has been discussed [7]. In the development of tiny-shaped particles, different precursors have been investigated [8]. Particles of different sizes were developed at various pulse rates, as discussed elsewhere [9]. There was a packing of tiny particles, developing nanoparticles and particles [10].

A study elsewhere [11] discussed a new pulse-based method for developing highly anisotropic gold particles. A deposited carbon film enhanced the field emission. The field emission is due to tiny graphitic grains, as discussed elsewhere [12]. The growth trend of grains and crystallites changes under different process conditions. A study elsewhere [13] discussed this. Under a suitable scheme, the atomic arrays of tiny particles can be converted into structures of smooth elements [14].

Structural evolutions in atoms of suitable elements are due to the execution of confined interstate electron dynamics [15]. By considering the silicon atom as a model, the phenomena of heat and photon energy were revealed [16]. X-ray diffraction is related to X-ray reflection and can be used to study new atomic structures [17]. The relation between force and energy at the electron level in a gaseous atom is different from that in a solid atom. A study elsewhere [18] discussed this.

A carbon atom with seven different states but a fixed number of electrons is discussed elsewhere [19]. Different element atoms are deposited to develop a hard coating, as discussed elsewhere [20]. The etching of photon energy in the deposition of carbon films has been discussed elsewhere [21]. The discussions of the studies cited in references 6 to 21 negate the concepts of ion formation. The atoms do not obey the concepts of ion formation. The existing study confirms this.

The study confirmed that an electric current is a photonic current. The bandgap is not due to the conventionally known conduction and valence bands. The existing study presents preliminary details on several well-known principles and phenomena with new insights. Photon propagation is unidirectional, bidirectional, or tri-directional if the gap between electronic states is one-dimensional, two-dimensional, or three-dimensional, respectively. This study not only advances science but also opens new chapters.

## 2.0. Experimental details/ Theoretical framework

The existing study does not contain any specific experimental details. The study develops a general understanding of both physical and chemical sciences. The existing study provides support for discussing the results in both ongoing and future research. The science presented here will also expand the discussion on different materials. In light of the existing study, the earlier discussed phenomena will align with the true scientific details.

The existing study provides an edge to discuss the results more insightfully. It also broadens knowledge in the engineering fields. The current study also helps to explain the results at the micron and bulk levels. To some extent, the study is relevant to medical sciences. In this context, it addresses both scientific and social impacts. In different areas of physical, chemical, environmental, and medical sciences, reliable discussions are an acute need. By referring to current investigations, such discussions are possible.

The electricity branch involves a lot of discussion about positive and negative charges. In light of the existing study, different ionization concepts require revision. The magnetism branch also requires revisions for several topics. To re-establish the different branches of physics and chemistry, revising several physical and chemical aspects is a prerequisite. The so-called plasma, which is a light glow in the existing study, also requires amendments in the discussion.

The current study especially covers the fields of atoms, solid-state physics, physical and chemical phenomena, microscopic analyses, sustainable and green sciences, energy and power sources, semiconductors, dynamics, materials physics and chemistry, medical physics, etc. This work is also useful for studying photon-matter interaction. The existing study rebuilds the science of semiconductor materials and inert gases. The study introduces a new era of science.

## 3.0. Models and Discussion

In a triangular-shaped tiny particle, arrays of atoms convert into the structures of smooth elements under the force of immersion. A study elsewhere [14] discussed this theoretically. A study elsewhere [22] discussed this experimentally. In those studies, the exertion of surface force tilts the laterally oriented electrons. The tilting of those electrons in their atoms is adjacent. A transitional-behavior atom studied the orientation of its left- and right-positioned electrons elsewhere [18]. These quoted studies do not indicate that atoms form ions.

The studies elsewhere [15, 16] discuss the confined electron dynamics in atoms of suitable elements. Those studies do not consider atoms with a positive or negative charge. A study elsewhere [19] discussed the non-confined electron dynamics in suitable carbon atoms. That study also discusses the partially confined electron dynamics in some other suitable carbon atoms. In those electron dynamics, carbon atoms do not indicate whether they lose or gain electrons.

The surface force does not significantly orient the central electrons of the atoms. In suitable solid atoms, the electrons of the inner rings can experience the surface force at a lower level. There is a minor tilting of those electrons. A solid atom should elongate when its electronic states behave plastically. The uniformly stretched energy knots should not recover in this case. In the elongation of a solid atom, the tilting electrons are along the east-west poles. The elongation of a solid atom does not relate to ion formation.

A solid atom has a distorted shape in its deformation. The stretched energy knots do not recover in a distorted atom. In this case, the orientation of the electrons can be partially adjacent and partially lateral. Electrons deal with the uneven forces. In the distorted shape of a solid atom, the energy knots clamping electrons alter their shapes. Electrons tilt non-uniformly in their atoms. Therefore, the mass of an atom does not remain conserved in its transitional behavior. In the original state, a solid atom keeps its conserved mass. Force and energy also remain conserved. Neither the conserved mass nor the non-conserved mass of the atoms indicates their ionization.

A hypothesized gold atom attains the re-crystallization state. Figure 1 (a) shows this. A gold atom can maintain a re-crystallization state at an electronically flat solution surface. To enter an electronically decreasing solution surface, a tiny particle leaves the flat surface. To elongate uniformly, that atom deals with the immersing force at an electronically decreasing solution surface. A study elsewhere [14] discussed both electronically flat and electronically decreasing solution surfaces.

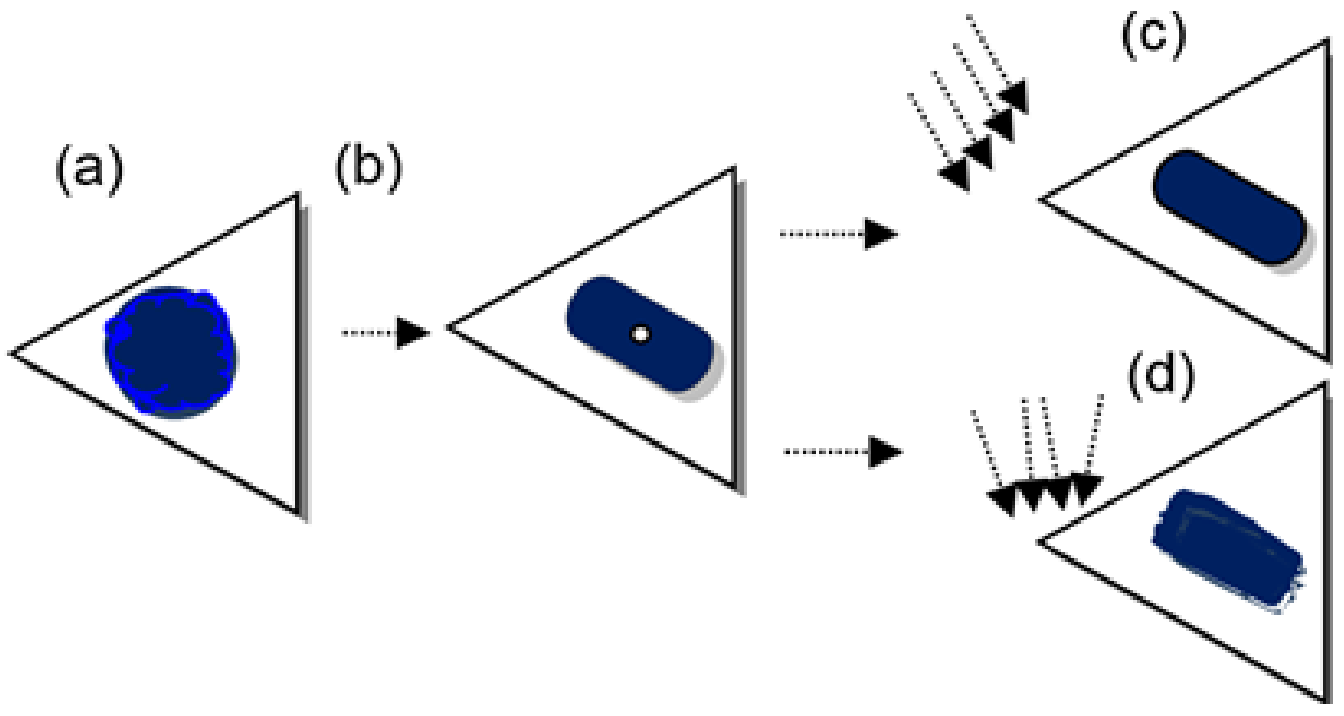

**Figure 1:** A single gold atom of a triangular-shaped tiny particle shows the (a) re-crystallized state at the flat solution surface, (b) elongation under immersing force, (c)

further elongation under the impingement of fixed-angle electron streams, and (d) deformation in the distorted shape under the impingement of different-angled electron streams. (Sketches drawn in estimation)

In the natural way of elongation of the gold atom, the electrons left of the center tilt south to the east, and the electrons right of the center tilt south to the west. (A study elsewhere [18] provides more detail on the electronic orientation of a hypothesized solid atom.) The dot in Figure 1 (b) denotes the center of the elongated atom. The electrons of the zeroth ring in an atom tilt marginally. Access to surface forces for the zeroth-ring electrons is nearly negligible. These factors indicate that the solid atoms do not ionize.

The impingement of the electron streams on a naturally elongated gold atom is at a fixed angle. In symbolic form, Figure 1 (c) shows this. As a result, a naturally elongated atom can further elongate. The electron streams impinge at a fixed angle to the naturally elongated gold atom, further stretching the energy knots. Thus, that atom undergoes further elongation. The impingement of the electron streams on a naturally elongated gold atom is not at a fixed angle. A naturally elongated gold atom does not undergo further elongation. That atom deforms into a distorted shape. In symbolic form, Figure 1 (d) shows this.

The lattice controls the orientation of electrons during the expansion of a solid atom. The transition energy changes the energy of the atomic lattice, as discussed elsewhere [23]. The expansion of the solid atom can take place throughout the lattice. The orientations of some electrons can reach the same level as in the recovery state of that atom. A solid atom can expand between the original and recovery states. That atom deals with the volumetric expansion. In this expansion, the surface force remains less influential. A solid atom expands linearly in its re-crystallized (or liquid) state. A separate study will discuss more details.

The gaseous atoms present different scientific insights. The lattice of a gaseous atom tightens under suitable conditions. Gaseous atoms contract rather than expand. The gaseous atoms undergo squeezing behavior. Gaseous atoms usually tighten their energy knots by squeezing rather than stretching. In gaseous atoms, the surface force also controls the orientation of electrons. During contraction, the lattice of a gaseous atom controls the orientation of electrons. Transitional energies of different forms can influence the atomic lattice. The contraction occurs throughout the atomic lattice.

In the contraction, the surface force usually remains less influential. The orientations of some electrons can reach the same level as in the recovery state of that atom. The orientations of some electrons can also change during the liquid state of that atom. There is a uniform contraction of gaseous atoms between the original and recovery states. In suitable elements, gaseous atoms can also contract between the re-crystallization and liquid states. These atomic behaviors also do not indicate the formation of ions.

In addition to the filled states, Figure 2 also shows an empty energy knot or unfilled state. An electron does not change its state while undergoing an infinitesimal displacement. There is no transfer of electrons to nearby unfilled states. The transition of a solid atom depends on the orientation of the electrons. The transition of a solid atom can result in its elongation (or deformation). The elongation (or deformation) of the solid atom indicates that it cannot form an ion.

**Figure 2:** (1) filled states and an unfilled state, (2) filled state electrons, and (3) unfilled state (or empty) energy knot

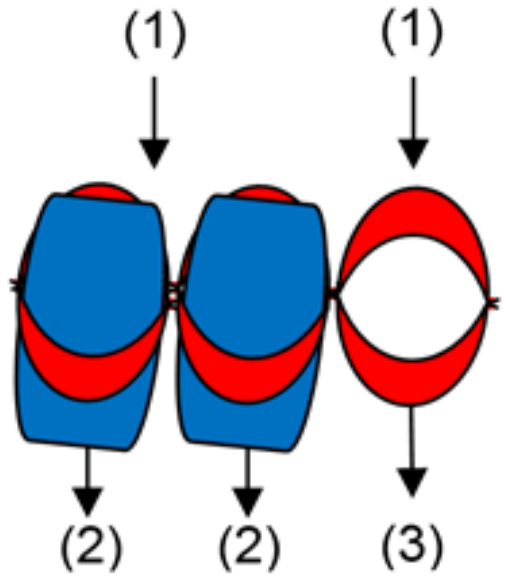

The erosion of solid atoms does not occur due to losing or gaining electrons. Instead, it is due to the extra stretching of the energy knots clamping the electrons. Squeezing of a gaseous atom is not due to losing or gaining an electron. Causing a gaseous atom to erupt is also not due to losing or gaining an electron. The splitting of the inert gas atom also does not refer to the ionization process.

A negative ion of the gold atom has the same number of electrons as a platinum atom. Moreover, a positive ion of a gold atom has the same number of electrons as a mercury atom. These findings also negate the earlier atomic structures in different elements. The atomic number does not remain the same. Because an atom loses or gains an electron. But the mass number for that atom remains the same, whether it loses an electron or gains an electron. Moreover, for both negative and positive ions, the mass of the gold atom is the same. However, both ions of the gold atom have different atomic numbers. Different atomic numbers for the same atom are not scientifically correct. It means atoms do not ionize. By losing an electron, one shell of the atom (with +1 valence) does not exist if considering the previously defined atomic structure. How is this possible? Therefore, atoms do not lose or gain electrons.

Inert atoms split under the field of a photonic current. The splitting is due to their inertness. It is also due to the photonic current field. A split inert gas atom neither loses an electron nor gains an electron. Kawai *et al*. [24] highlighted the role of classical van der Waals interactions under the limit of an isolated atomic model. The binding of inert gas atoms can be due to other factors. The attractive force is due to the induced dipoles. The attractive forces are the van der Waals forces, as discussed elsewhere [25]. They are also dispersion forces. Binding atoms requires some additional steps to discuss. It is not only a force that binds the atoms. Regarding the binding of atoms, studies in references 10 and 14 provide further details.

Label (1) in Figure 3 shows the argon atom. The splitting of argon atoms is due to the field of the photons. The splitting of argon atoms is usually the case in a pulse-based process. Inert gas atoms convert into electron streams. The argon atoms split under the excessive field of the photons. Label (2) shows the splitting path of the argon atom. Label (3) in Figure 3 shows the tube bottom. It deals with the flowing inert gas atoms and the photonic field. Label (4) shows the electron streams carrying the photons. Label (5) in Figure 3 shows the traveling photons. They travel without following the electron streams.

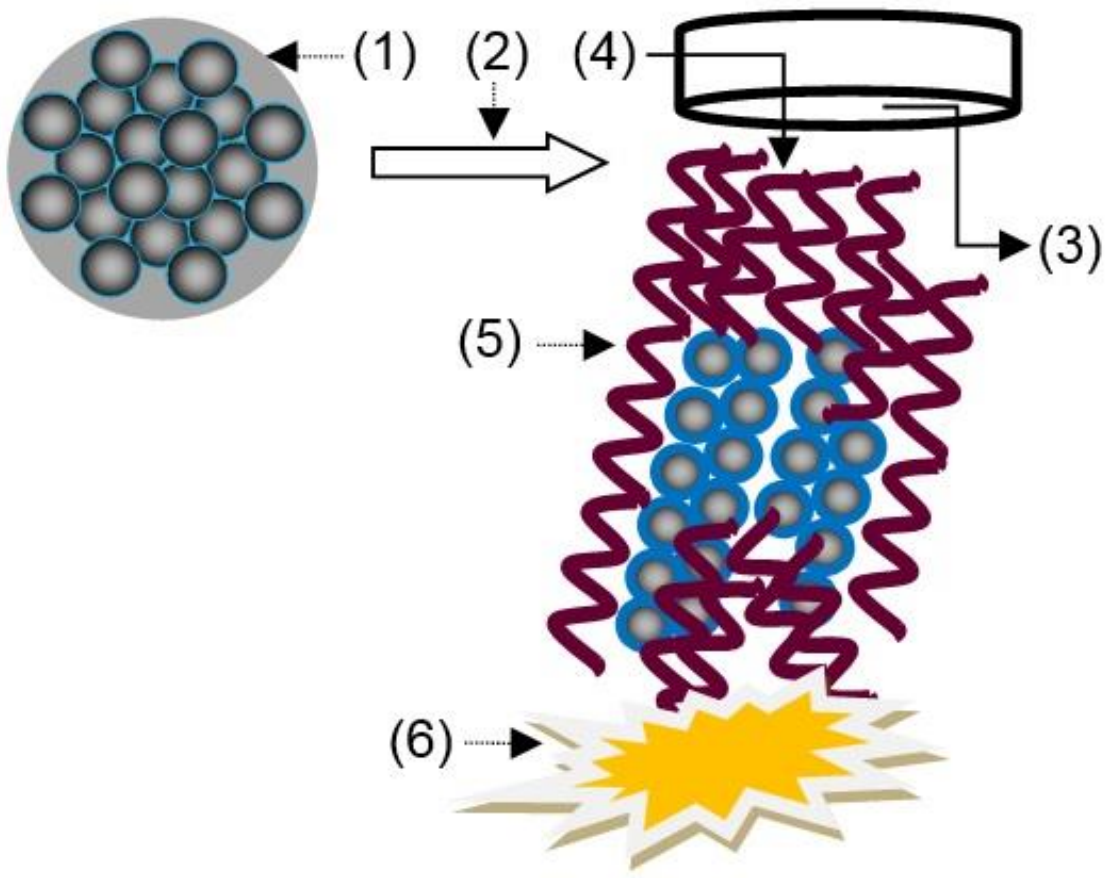

**Figure 3:** (1) Argon atom, (2) splitting argon atom just below the tube, (3) tube handling flowing inert gas atoms and an excessive current field, (4) electron streams carrying photons, (5) direct entrance of photons to the air medium, and (6) glow of light

In solution processing, the photons can also enter the solution. When photons leave the splitting inert gas atoms, they enter the solution or air medium. Their characteristics become apparent. Light appears at the point of confinement of two fields. Label (6) in Figure 3 denotes a light glow. A light glow is below the bottom of the tube. The light is due to the confinement of two field forces. The traveling photons and the air medium confine their field forces. As a result, two forces of different fields interact to show lighting.

Photons propagate in the gaps between electronic states. Inert gas atoms behave inertly because they cannot undergo electron dynamics. Inert gas atoms also do not undergo transitions. The photons of the current split the inert gas atoms. The splitting of the inert gas atoms occurs due to the propagation of the photons between the gaps of their electronic states. As a result, the photons enter the air medium. The wavelength of the photons can reach the visible range. The local confinement of the forces is the cause of lightning behavior. Those forces belong to different sources. A light spot or glow usually indicates an orange color. It also relates to the so-called plasma.

The dynamics of an electron generate the photon characteristics of the current. The dynamics of an electron are the case in a silicon cell or other similar device. A study elsewhere [16] discussed this. The entry of a photon into the grid is due to the suitable fabrication process of the solar cell. Photons propagate through the electronic state gaps of atoms to reach the terminals of solar cells.

The flow of electrons or the concept of ion formation does not correctly explain the science. The photonic current refers to the propagation of photons through a suitable medium. The photons propagate through the gap between the electronic states of a material. The photons propagate through the electronic state gap. The electronic state gap, or the gap between the electronic states, is the same thing. It can also refer to a photonic bandgap. Therefore, a current is not due to the flowing electrons (or charged particles). The propagation of the photons from a suitable wire defines the photonic current.

In different microscopes, the image of a specimen captured by the Camera appears on the computer or desktop screen. Before capturing an image, the featured photons released from the built-in component of a microscope resolve the surface (of interest). The resolution of the surface in that image is up to a suitable magnification. The numeric value represents the magnification. It is usually available at the image corner. A scale bar also denotes that magnification. In addition to the type of microscope, the image magnification depends on the selected scale.

The power of a high-resolution transmission microscope is at the subatomic level. However, this is not the case with the field emission scanning microscope. The photons of the current characteristics can melt the sample under investigation. Thus, the application of a transmission microscope already has full resolving power. In a high-resolution transmission microscope image, the approximate width of the elongated atom was 0.05 nm, as discussed elsewhere [9]. The resolution of that image was due to the transmitted photons. The release of those featured photons was from the built-in component of that microscope.

In the cathodic arc evaporation technique, the arc ignition is due to the high field of the featured photons (or photonic current). The metallic atoms are deposited on the substrate to develop a hard coating, as discussed elsewhere [20]. Before their deposition on the substrate, the atoms detached from the metallic target. A suitable field of the random arc detached the atoms from the metallic target. Thus, an electric current is a photonic current.

In the so-called electrolysis process, the flow of anions toward the anode is not due to gaining electrons. Both the energy and force of photons cause the atom to dissociate from the precursor or compound. There is also a need to revise the science of lithium-ion beam technology. In a focused ion beam, sample etching occurs due to the suitable population of featured photons rather than ions [26].

This study clearly reiterates that neither electrons nor charged particles can flow. Therefore, the electric current is an incomprehensible phenomenon. The photonic current is a comprehensible phenomenon. A photon with current characteristics has a width within the electronic state gap, as shown in Figure 4. Label (1) in Figure 4 denotes the width of a unit photon. Label (2) in Figure 4 indicates the electronic state gap, or the gap between electronic states.

**Figure 4:** Unit photon having characteristics of the current (1) width and (2) electronic state gap

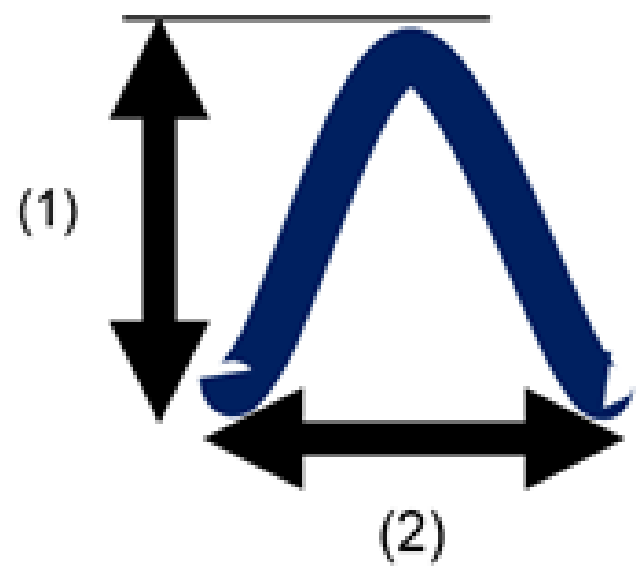

The wavelength of a unit photon is equal to the gap between the filled and nearby unfilled states of a silicon atom. The filled and nearby unfilled states belong to the outer ring. A study elsewhere [16] discussed this. An aligned electronic state gap exists between the electronic states. The distance (or gap) between the start and end points of the unit photon equals the width of that unit photon. Figure 4 shows this. In crystalline and electronic materials, the photons propagate through the gaps between electronic states. The propagating photons carry the force and energy to the other end of the material.

A material depends on its processing conditions in attaining a consistent gap between electronic states. The processing conditions of a material are in addition to the properties of participating atoms. Depending on the crystallinity and other features of a processed or synthesized material, the gap between the electronic states of the bound atoms changes. As a result, the number of propagating photons from a material varies. Gaps between electronic states can exist in one-dimensional, two-dimensional, and three-dimensional systems.

The photon propagation in Figure 5 is in different directions. If there is a one-dimensional gap between the electronic states, there is also unidirectional propagation of the photons. Figure 5 (a) shows this. If a material has a two-dimensional gap between the electronic states, the photons propagate in two directions. Figure 5 (b) shows this. If a material has a three-dimensional gap between the electronic states, the photons propagate in three directions. Figure 5 (c) shows this. Figures 5 (a), (b), and (c) also show the input point of the photonic current field.

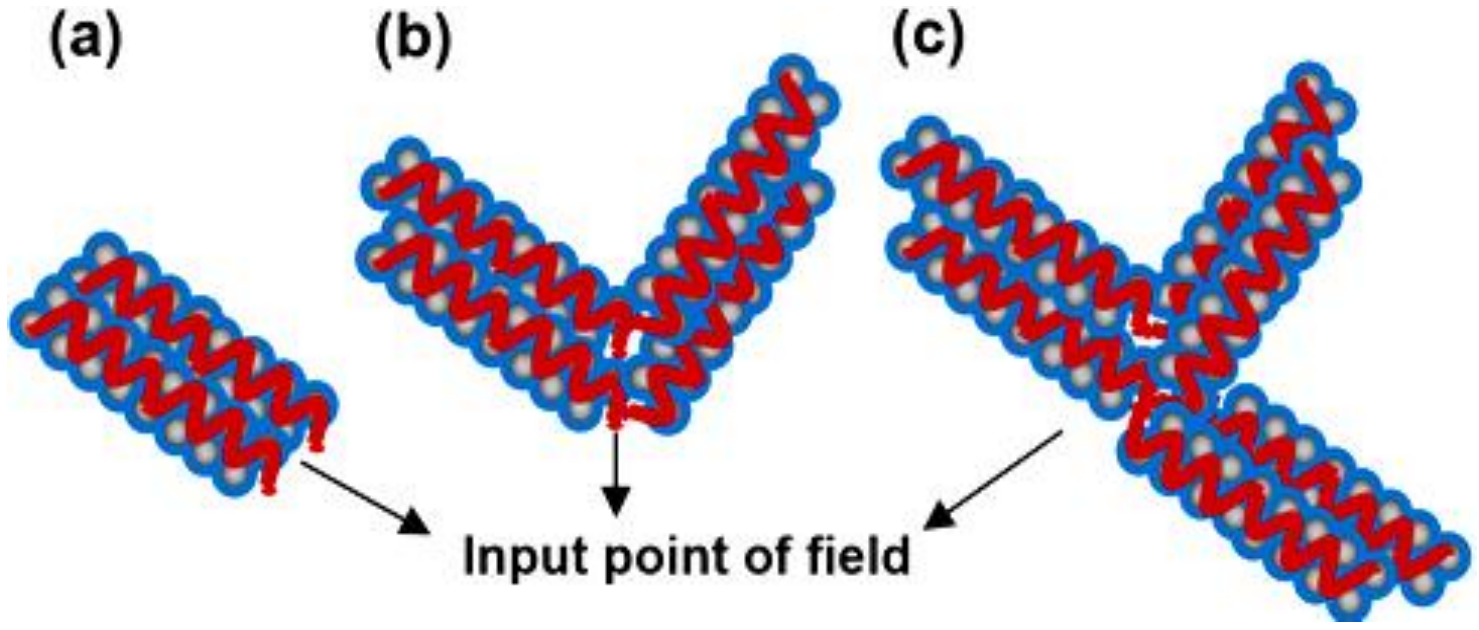

**Figure 5:** Propagating photons in the photonic bandgap (or gap between the electronic states) (a) unidirectional, (b) bidirectional, and (c) tri-directional

In a material, the gap between electronic states can occur in four-, five-, or six-dimensions. As a result, photons can propagate in tetra-, penta-, or hexa-directions, respectively. Hence, many studies are required to understand the bandgap in materials with different structures. The gap between the electronic states refers to a photonic bandgap. In tiny grains of a carbon film, photons propagate through the gaps between electronic states [12]. That carbon film functions to emit an enhanced field. As discussed elsewhere [16], a unit photon is a slice of an overt photon.

Figure 6 shows three 'unit photons'. Each unit photon has a different color. Figure 6 also shows the perturbed-state electrons. Circles filled with red color in Figure 6 denote the perturbed-state electrons. In the case of the perturbed-state electrons of the bound atoms, the "unit photons' deal with uneven interactions. The misaligned electronic states do not permit the 'unit photons' to propagate. Photons do not propagate in a material with misaligned electronic state gaps. Figure 6 shows this. This kind of material exhibits insulating behavior.

**Figure 6:** Unit photons converting into heat energy upon interacting with the misaligned electronic states

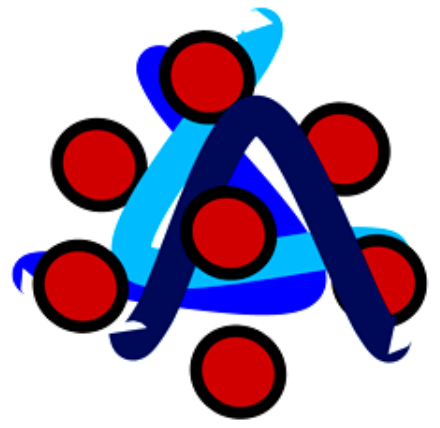

In some materials, only some portions have the electronic alignment. Those small portions have aligned electronic state gaps. The propagation of the photons is only from those portions. The short-circuiting is due to the misaligned gaps between the electronic states of the metallic-atom wire. In a broken photon, the energy and force elements are locally isolated. Photons also convert into heat energy. In this case, photons do not propagate through the gaps between the aligned electronic states of a material.

Photons interact with the electronic tips in different ways. A study elsewhere [16] discussed some details about this phenomenon. When the electronic state gaps are uniformly aligned, photons propagate without breaking. Propagating photons do not convert into heat energy. In a material where atoms keep aligned electronic state gaps, the propagation of the photons is without any loss. It was a “conductor” in earlier studies.

In a crystalline material, the propagating photons do not break into pieces. By preserving both the nature of energy and the nature of force, photons propagate to the output end. In titanium nitride coatings, the deposition of gaseous atoms and solid atoms results in insulating behavior [20]. A band gap for electrons to flow does not exist in any material. So, there is no flow of electrons or charged particles. A photon propagates in the gap between the electronic states, or the photonic bandgap.

A study elsewhere [23] discussed the significance of an electronic orientation in the atoms of suitable elements. In addition to entropy and geometry, ongoing research efforts should also consider dynamics [27]. By moving optical tweezers in a real-time control system, the geometry of cold neutral atoms for quantum engineering was prepared [28]. A study elsewhere [29] discusses the preparation of regular arrays of cold atoms by moving optical tweezers in a real-time control system. Research and development in the energy sector require new approaches. Research in the energy sector will reduce losses. It will help individuals build a green environment.

In the context of a new atomic structure, there is a need to explore photon or light-matter interaction. The existing study validates that flowing (and splitting) inert gas atoms can cause photons to glow. The photons light up by extending their wavelength.

Inert gas atoms can clarify the electron-photon phenomena. Studies of inert gas atoms can be critical in the medical and biological sciences.

All these glitters need not only gold but also titanium nitride. A study elsewhere [30] discussed this. Four titanium atoms maintain a nitrogen atom at the interstitial position. In this way, a unit cell in titanium nitride forms. The unit cell of titanium nitride gives almost the same number of states as a gold atom. It might be the reason for the glittering (golden color) of the titanium nitride coating. There is a need to study the glittering of other elements and compounds under suitable conditions.

The studies discussed elsewhere [31-50] mainly focused on the conduction band, ionization, or electric current. The existing study does not agree with those cited studies. The conduction (or valence) band is no longer a requirement of modern science. As discussed above, atoms also do not form ions. Instead, they modify their structures in different ways. A study elsewhere [51] discussed different modification behaviors of the atoms. As discussed above, an electric current is now a photonic current.

In the elongation of a solid atom, the supplied energy uniformly aligns the perturbed-state electrons. The competing forces exerted at the electron level are even. A crystalline material keeps a uniform atom-level modification. In the distorted atom, the perturbed states do not align their electrons. The atom-level modifications in an amorphous material are not uniform. Many scientific phenomena in the developed processes, devices, and instrumental techniques await their revision.

## 4.0. Conclusions

Atoms do not form ions. Different atomic behaviors indicate this. In its original state, the mass of an atom remains conserved. In the transition state, an atom does not possess a conserved mass. In introducing the transitional behavior of an atom, the surface force is vital. From its atomic center, left-sided electrons tilt from south to east under the exertion of the surface force. The tilting of the right-positioned electrons is from south to west under the surface force. It is not feasible that an atom loses or gains an electron.

Due to the impingement of the electron streams at different angles, an earlier elongated atom can deform into a distorted shape. In a distorted atom, the stretching of energy knots that clamp electrons is non-uniform. When the impingement of the electron streams has a fixed angle, an elongated atom elongates further. A solid can erode during extended elongation. Atoms of gaseous elements contract. They tighten the energy knots, clamping electrons. Squeezing occurs. A gaseous atom can erupt if the squeezing continues. In different elements, atoms do not form ions.

When argon gas flows through a suitable apparatus, the inert gas atoms split under the field of photonic current. (The gas of the suitable atoms can flow from the bottom of a tube or from a suitable setup.) By decreasing their wavelength, photons travel in the air. They light up in the visible range. Two different-natured forces confine their fields. As a result, a glow of light appears. A force of photons mixes with the force of the air.

The interstate electron dynamics of a silicon atom generate wave-like photons, as discussed elsewhere [16]. Photons enter the grid. They function as a photonic current. To resolve the surface of interest, different microscopes emit photons with suitable characteristics. On the computer screen, there is an image of the under-study sample surface. The photons propagate from an aligned electronic state gap or a photonic bandgap rather than from the conventionally known bandgap.

A current through a metallic wire is a photonic current, as it refers to the propagation of photons. The flow of electrons (so-called an electric current) is extensively studied; this is not the case. If the gap between the electronic states is one-dimensional, the propagation is unidirectional. If the gap between electronic states is two-dimensional, the photons propagate in two directions. The photons propagate in three directions when the gap between electronic states is three-dimensional.

An electric current is the photonic current. Several well-known principles and phenomena agree on this. A highly crystalline material ideally supports photon propagation. In an amorphous material, there is no photonic bandgap. The existing study enables one to understand basic and applied science, various physical and chemical phenomena, modern physics, microscopic principles, materials science and engineering. The existing study truly opens multiple avenues for scientific research.

**Funding declaration:**

There is no funding.

**Acknowledgments**:

The author acknowledges the support of whoever he learned from at any stage.

**Data availability statement:**

This study does not contain any data.

**Conflicts of interest:**

The author declares no conflict of interest.

## Bibliographic detail:

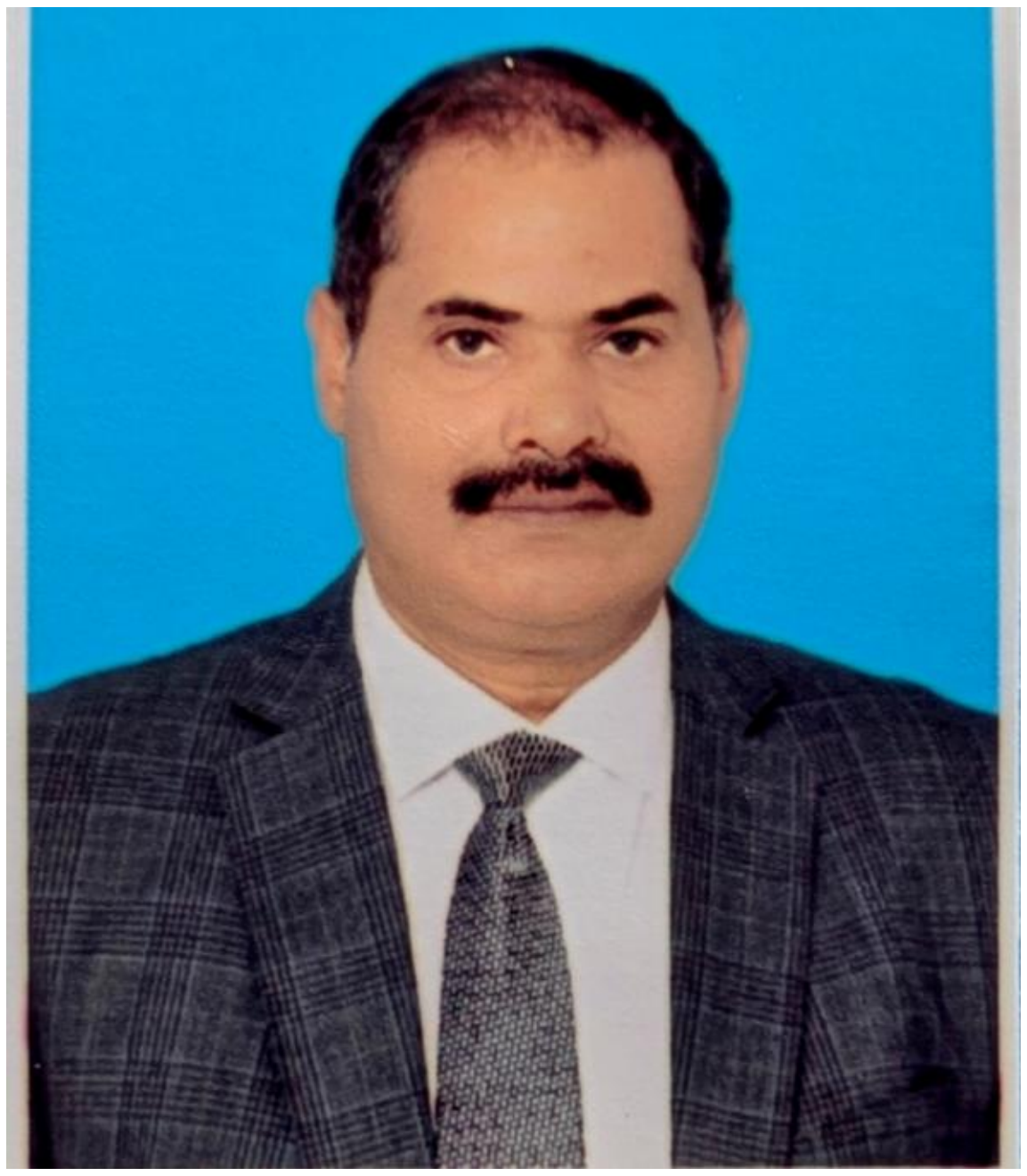

In the year 1996, **Mubarak Ali** obtained a B.Sc. degree in Physics and Mathematics. It was from the University of the Punjab. Mubarak Ali earned the M.Sc. degree in Materials Science in 1998. Mubarak Ali completed the MSc degree with distinction from the Bahauddin Zakariya University, Multan. He completed the MSc thesis at Quaid-i-Azam University, Islamabad. From Universiti Teknologi Malaysia, Mubarak Ali earned a PhD degree in Mechanical Engineering (Major in Materials Engineering). The Malaysian government awarded a PhD scholarship under the Malaysian Technical Cooperation Programme (MTCP) in 2004-07. In 2010, the Scientific and Technological Research Council of Turkey (TÜBİTAK) funded the postdoctoral research under the foreign fellowship programme. Dr Mubarak Ali obtained the first postdoc position in advanced surface technologies from the Department of Metallurgical and Materials Engineering at Istanbul Technical University. Dr Mubarak completed another postdoc in nanoscience and nanotechnology from Tamkang University (Physics Department), Taipei, in 2013-2014. The National Science Council, now the Ministry of Science and Technology, Taiwan, sponsored his second postdoctoral studies. He worked as an Assistant Professor on the tenure track at COMSATS University Islamabad from May 2008 to June 2018 (previously known as the COMSATS Institute of Information Technology). His new position is in process. He also worked as an assistant director and deputy director in the Pakistan Council of Renewable Energy Technologies, Ministry of Science and Technology, Islamabad, from January 2000 to May 2008. The Institute for Materials Research at Tohoku University, Japan, invited Dr Mubarak to deliver a scientific talk. He presented scientific research at many conferences and in many countries. The main research areas include basic science, physics, materials science, surface science, surface engineering, materials chemistry, physical chemistry, energy science, nanoscience, and sustainability. Dr Mubarak also won a merit scholarship for PhD studies from the Higher Education Commission, Pakistan. Due to another position, he did not obtain this opportunity. In the year 2000, Mubarak earned a diploma in English language from the National University of Modern Languages, Islamabad. From the same institution, Dr Mubarak Ali completed the Japanese language certificate in 2001. Dr Mubarak Ali is the author of several articles available at the links https://www.researchgate.net/profile/Mubarak_Ali5 & https://scholar.google.com.pk/citations?hl=en&user=UYjvhDwAAAAJ